\documentclass[
reprint,
superscriptaddress,
amsmath,
amssymb,
aps,
prb,
longbibliography,
noeprint
]{revtex4-2}

\usepackage{epsfig}
\usepackage{dcolumn}
\usepackage{bm}
\usepackage{color}
\usepackage{graphicx}
\usepackage{bbold}

\begin{document}

\title{Quantum phase transition in small-size 1d and 2d Josephson junction arrays: analysis of the experiments within the interacting plasmons picture}

\author{Samuel Feldman} 
\affiliation {Department of Physics and Astronomy, University of Utah, Salt Lake City 84093, USA}
\author{Andrey Rogachev} 
\affiliation {Department of Physics and Astronomy, University of Utah, Salt Lake City 84093, USA}

\date{\today}

\begin{abstract} 
Theoretically, Josephson junction (JJ) arrays can exhibit either a superconducting or insulating state, separated by a quantum phase transition (QPT). In this work, we analyzed published data on QPTs in three one-dimensional arrays and two two-dimensional arrays using a recently developed phenomenological model of QPTs. The model is based on the insight that the scaled experimental data depend in a universal way on two characteristic length scales of the system: the microscopic length scale $L_0$ from which the renormalization group flow starts, and the dephasing length, $L_{\varphi}(T)$ as given by the distance travelled by system-specific elementary excitations over the Planckian time. Our analysis reveals that the data for all five arrays (both 1D and 2D) can be quantitatively and self-consistently explained within the framework of interacting superconducting plasmons. In this picture, $L_{\varphi}=v_p\hbar/k_B T$, and $L_0 \approx \Lambda$, where $v_p$ is the speed of the plasmons and $\Lambda$ is the Coulomb screening length of the Cooper pairs. We also observe that, in 1D arrays, the transition is significantly shifted towards the insulating side compared to the predictions of the sine-Gordon model. Finally, we discuss similarities and differences with recent microwave studies of extremely long JJ chains, as well as with the pair-breaking QPT observed in superconducting nanowires and films.
\end {abstract}

\maketitle

\noindent \textbf{1. Introduction}

Josephson junction (JJ) arrays are a rich subject of study, both for their applications in superconducting devices and for the interesting physics they exhibit. These arrays can introduce a very high inductance to a superconducting circuit while maintaining negligible dissipative loss \cite{BellSadovskyy,Masluk}, effectively behaving as “superinductors.” The Fluxonium qubit, for instance, relies on this inductance to isolate the qubit from charge fluctuations \cite{ManuFluxonium}. Additionally, other proposed qubit designs utilize a JJ array as the active element in their architecture\cite{KitaevQubit}.

Josephson junction arrays also serve as a relatively simple and well-controlled platform for exploring the so-called "phase-only" superconductor-insulator transition (SIT). However, decades of research have shown that this apparent simplicity can be quite deceptive. 

Theoretical understanding of such a transition goes back to the work of Bradley and Doniach mapping the 1d chain onto a 2D XY model (with imaginary time providing the second dimension) \cite{BradleyDoniach}. They showed that the state of the chain is determined by three energy scales associated with three electric parameters: 1) Charging energy between the islands $E_1$, which depends on the corresponding capacitance as $E_1=\left(2e\right)^2/2C_1$; 2) Josephson coupling energy $E_J$, characterized by the junction inductance as $E_J=\hbar^2/\left(2e\right)^2L_J$; and 3) Capacitive charging energy to the ground $E_0=\left(2e\right)^2/2C_0$. (There is also a fourth parameter, the shunting resistance of the junction, $R_{sh}$, which could drive a dissipative QPT \cite{BobbertSchon}. However, for the systems discussed in this paper, it is very large and thus not relevant). When analyzing a 1D JJ array, additional parameters are introduced: the screening length $\Lambda=\sqrt{C_1/C_0}$ and dimensionless ratios $K_0=\sqrt{E_J/E_0}$ and  $K_1=\sqrt{E_J/E_1}$ which determine superconducting correlations for $\Lambda >1$ and $\Lambda <1$, respectively. 

In the limit of the very long arrays, Bradley and Doniach predicted that, if $C_0=0$, the array couples to itself and is always insulating. If, however, the two capacitances are comparable, the quantum state of the chain is determined solely by the competition between $E_J$ and $E_0$. When $E_J$ dominates, the transport is similar to a superconductor, and when $E_0$ dominates, the system behaves as an insulator.

The theoretical literature on 1d JJ arrays is extensive. Recent studies include a mapping onto the sine-Gordon model in order to better investigate the dynamics of each phase \cite{HermonSoliton,HomfieldSoliton}. The model allows to treat the array as a Luttinger liquid, where parameter $K_0$ determines the phase transition \cite{GlazmanLarkin,FazioVortex}. In Ref. \cite{MirlinSIT}, the authors extended the model by incorporating fluctuations from quantum phase slips (QPS) and the effect of random offset charges. Using numerical renormalization group analysis, they predicted how the transition is influenced by temperature, array length, and disorder, specifically for short-range $(C_1\ll C_0)$  and long-range $(C_1 \gg C_0)$ interactions. A zero-temperature phase diagram for arbitrary interaction range has been determined in the study of disorder-free chains with an imaginary-time path integral quantum Monte Carlo algorithm \cite{BaskoQMC}. 

Experimentally, the existence of the superconducting and insulating states was established in several classical works on relatively short arrays of SQUIDs (with $N$=63–255) \cite{HavilandArrays,HavilandReview}. However, despite numerous subsequent observations of SITs, \cite{HavilandQPS,MiyazakiQPT} there has been no agreement with theory on the location of these transitions. Another point of concern is the behavior of the arrays at the lowest temperatures: many curves that are reported to be in the insulating state exhibit re-entrant behavior, where the resistance decreases as the temperature is lowered \cite{HavilandArrays}. This observation raises the question of whether the transition from the superconducting state occurs through a true quantum phase transition (QPT) or merely as a crossover.

The SIT has also been studied in 2D arrays, although less extensively. Both the superconducting and insulating states are predicted to undergo a Berezinskii-Kosterlitz-Thouless (BKT) transition at low enough temperatures. The key difference is that in the superconducting state, vortices become localized, while in the insulating state, charge localization occurs. The critical point of the transition is expected to occur at the self-duality point, where charge and phase are interchangeable, specifically when $E_J/E_1 = 2/\pi^2$ \cite{FazioVortex}.

The 2D behavior was studied experimentally in \cite{Haviland2d,ZantMooij2d}. While a SIT mediated by $E_J/E_1$ has been reported, there was a flattening of the curves at low temperatures and inconsistencies in the values of the critical exponents. 
\begin{figure*}[tbph]
\centering
 \includegraphics[width= 1.0\textwidth]{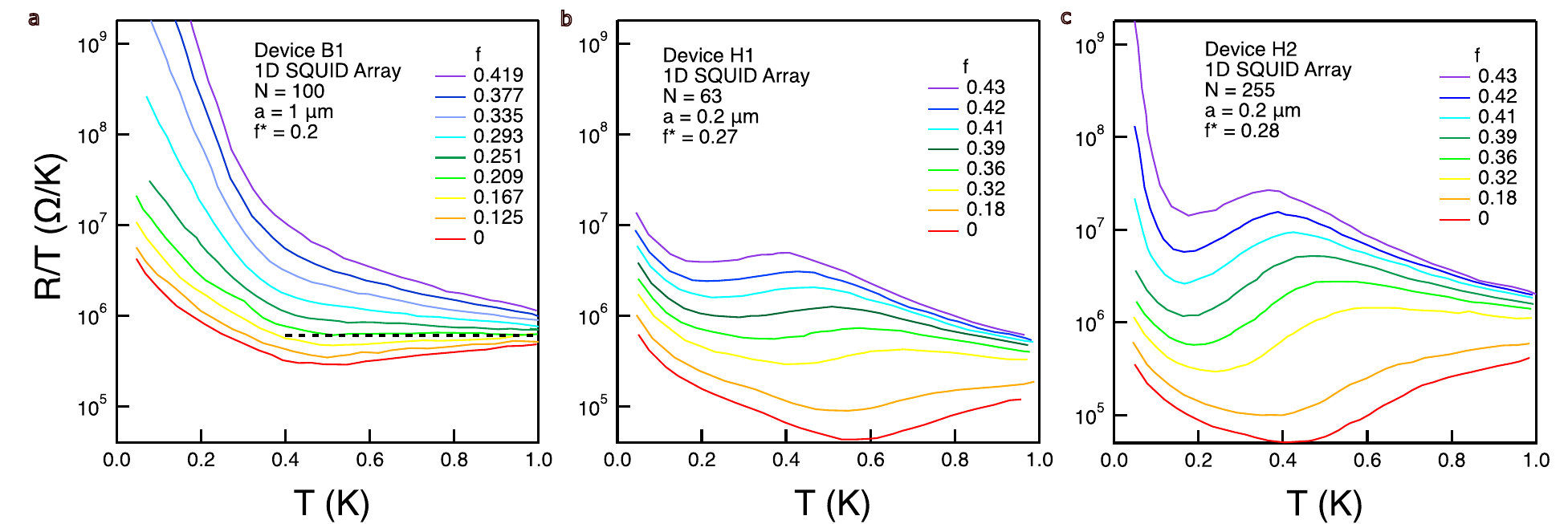}
 \caption{\textbf{Scaled resistance versus temperature for 1D Josephson junction arrays.} 
\textbf{(a)} The scaling plot for array B1 studied in \cite{KuoChen}, the horizontal line indicates the separatrix between the insulating and superconducting regimes for the data at $T>$0.5 K. \textbf{(b,c)} The scaling plots for arrays H1 and H2 studied in \cite{HavilandArrays}.}
\end{figure*}
The recent resurgence of interest in Josephson junction arrays is driven by their numerous applications in quantum computing, as well as the development of new methods for exploring the superconductor-insulator transition. Advancements in fabrication techniques have enabled the production of arrays with tens of thousands of junctions, while microwave measurement techniques now allow for the direct observation of plasmons in these arrays \cite{KuzminManucharyan}. Surprisingly, this study found high-frequency excitations propagating without dissipation, even in arrays that were predicted to be deep in the insulating regime. A similar phenomenon was observed in \cite{HigginbothamMelting}, where it was suggested that the insulating state melts into a state exhibiting local short-range superconductivity. This crossover from global to local behavior was studied using microwave spectroscopy. However, the corresponding behavior in direct current (DC) remains unclear. On a broader scale, it is also uncertain whether the transition from the superconducting to insulating states occurs as a continuous quantum phase transition, and if so, what the microscopic mechanism behind this transition is. 

In this paper, we analyze the existing DC data of the arrays using a newly developed empirical model of QPTs \cite{RogScale}. Unlike the standard finite-size scaling analysis, which provides just the critical exponents (such as correlation length 
exponent $\nu$ and dynamical exponent $z$), our model gives insight into the microscopic physics governing the transition. It is applicable in the quantum critical regime where dephasing length $(L_\varphi \sim 1/T^{1/z})$ is smaller than the correlation length $(\xi \sim \left|y-y_c\right|^{-\nu})$. In this regime, the resistivity can be approximated as
\begin{equation}
R\left(T\right)=\frac{\hbar L_\varphi^{d-2}}{g_ce^2}\exp{\left(\frac{y-y_c}{y_c}\left(\frac{L_\varphi\left(T\right)}{L_0}\right)^{1/\nu}\right)}
\end{equation}
In addition to the exponential approximation, the second defining feature of the model is the inclusion of $L_0$, the microscopic scale of the transition. This quantity is generally known and represents a minimum scale at which the quantum phases can exist and from which the renormalization group flow starts.

This scaling behavior described by Eq. 1 is remarkably universal. By taking the dephasing length as the distance a non-interacting semiclassical carrier (or excitation) travels over the Planckian time, $\tau_p=\hbar/k_BT$, $L_0$ has been found to correspond closely to an expected minimum length scale in various systems. This includes mean free path in doped semiconductors, lattice constant in cold atomic gases and moiré superlattices \cite{RogScale}, magnetic length in quantum Hall systems \cite{RogPlanck}, and coherence length in superconducting MoGe nanowires, a variety of superconducting films, and La$_{1.92}$Sr$_{0.08}$CuO$_4$ \cite{RogPB}.

In this paper, we will use this model to analyze the DC resistance data for three SQUID arrays studied in Refs. \cite{HavilandArrays,KuoChen}, as well as square 2d arrays studied in \cite{ZantMooij2d,Nakamura2d}. Let's start with considering 1d arrays. 

\smallbreak
\noindent \textbf{2. QPT in 1D SQUID chains}

The QPT in 1d SQUID arrays was tuned by changing the magnetic field piercing the SQUID loops. In the first step of the analysis, we have traced $R\left(T,B\right)$ data for arrays H1 (N = 63) and H2 (N = 255) from Fig. 3 of \cite{HavilandArrays} and for array B1 (N=100) from Fig. 3b of \cite{KuoChen} and analyzed them by comparing the data with the standard equation of finite size scaling
\begin{equation}
R\left(T\right)=\frac{\hbar}{e^2}T^\frac{-\left(d-2\right)}{z}\Phi_R\left(\frac{y-y_c}{y_c}\frac{1}{T^{1/z\nu}}\right)
\end{equation}

To compare data with Eq. 2, we need to take account for the temperature dependence in the prefactor of the exponent. Because we are studying a 1d system, in order to collapse all of the data onto a single curve, the measured resistance values must be divided by $T^{1/z}$. From our model, as explained below, we expect $z=1$. The study of the 1d array also predicts that at the critical point, $y=y_c$, resistance varies linearly with temperature \cite{MirlinSIT}. Following these predictions, in Fig. 1 we show the normalized resistance $R(T)/T$ versus temperature for three analyzed SQUID arrays. The normalized flux threading the SQUID is indicated for each curve.  The same vertical scale was chosen for all three panels to allow for one-to-one comparison between the arrays.

We can see from the figure that, when plotted in this way, the data at $T>0.5\ K$ display a fan-like behavior with a flat separatrix. Below $0.5\ K$, the resistance of the curves deviates towards insulating behavior. The reason for these deviations will be explored later, but in the interest of performing the scaling analysis we will (following Kuo and Chen \cite{KuoChen}) only look at the data above $0.5\ K$. 

While the transition in the SQUID array is physically driven by an applied magnetic field, this is actually a proxy for a more fundamental critical parameter. Theoretically, the QPT in a long array is driven by $K_0=\sqrt{E_J/E_0}$. Let us now quantitatively define this and other relevant parameters of the arrays. The unit cell is 0.2 $\mu$m in arrays H1 and H2 and 1 $\mu$m in array B1; we use these as length scales so in the text below “per length” and “per unit cell” is the same. The studied JJ chain is a 1d array of SQUIDs, so a unit cell has two Josephson junctions in parallel. The Josephson energy per unit cell varies with magnetic field as $E_J=E_j^0\cos{\left(\pi f\right)}$, where $f=AB/\Phi_0$ is the normalized flux, $A$ is the area of the SQUID loop, B is the magnetic field, $\Phi_0$ is the flux quantum. The zero field Josephson energy per unit cell, $E_J^0$, was reported to be 130 $\mu$eV for device B1 \cite{KuoChen} and 142 $\mu$eV for devices H1 and H2 \cite{HavilandArrays}. Then, the inductance per unit cell can be found as $L_J=\hbar^2/\left(2e\right)^2E_J$.

We refer to the capacitance between neighboring islands as $C_1$ and its value was determined in Refs. \cite{HavilandArrays,KuoChen}. The capacitance to ground, denoted as $C_0$, is provided for H1 and H2, however we had to estimate its value for B1. $C_0$ there comes from two main contributions. Array B1 was fabricated with a coplanar gate electrode, which had width 0.5 $\mu$m and was located at a distance of 1 $\mu$m from the array \cite{KuoPriv}. It gives a capacitance $C_{0g}\approx0.18$ fF.  The sample was fabricated on a 670 $\mu$m-thick Si wafer, which during the measurements was placed on a copper plate \cite{KuoPriv}. Capacitance to this plate estimated via the microstrip equation gives the contribution  $C_{0m}\approx0.086$ fF. Combining these two contributions gives a total $C_0\approx0.27$ fF. 

With thus determined values of $C_0$ and $E_J$, the normalized flux for each curve in Fig. 1 was converted into the corresponding parameter $K_0$. As mentioned earlier, there are two different potential critical parameters for this transition $K_0$ and $K_1$; which one is proper will be discussed in the analysis. However, both have the same dependence on applied magnetic field, $B$. Because $y$ is normalized in Eq. (2), the choice of $K_0$ or $K_1$ does not affect the analysis.

According to Eq. (2), each of the resistance curves for a single device should collapse onto a single curve when plotted against scaled $K_0$. In this analysis two parameters, the correlation length exponent $\nu$ and the coupling constant at the critical point of the transition $K_0^\ast$, are allowed to vary. Fig. 2a shows the $R/T$ data for device B1 plotted in log-linear axis against the scaled variable $\left(K_0-K_0^\ast\right)/K_0T^{1/z\nu}$. The inset shows the corresponding standard log-log plots. The scaling analysis according to Eq. 2 was already performed in \cite{KuoChen}. The parameters found in the original paper were $f_c\approx0.2$, $z\approx1$ and $\nu\approx0.45$. From our own analysis, we found the same critical flux and dynamical exponent, however, the correlation length exponent is different, $\nu\approx0.65$. The discrepancy in $\nu$ occurs because in the original paper, the authors apparently tried to get the best collapse on a log-log scale (see inset) over a broad range of parameters, while our goal is to get a compact set of data across the transition. We do see some deviation from the expected scaling behavior for curves with $K_0=0.17,0.21$ in Fig. 2b. This is most likely caused by these curves being too far from the critical point for the scaling to be valid.
\begin{figure*}[tbph]
\centering
 \includegraphics[width= \textwidth]{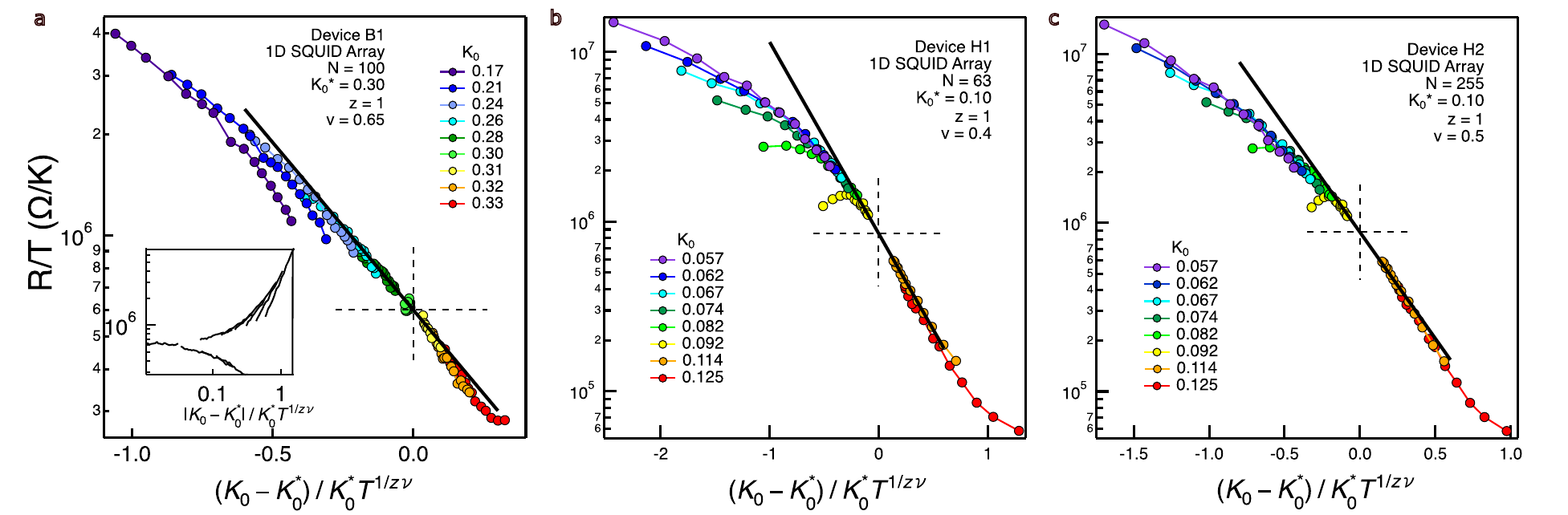}
 \caption{\textbf{Scaling analysis of 1D arrays.}
\textbf{(a)} Resistance vs. scaled coupling energy for array B1. The black solid line is the exponential fit to the data across the critical point. The inset shows the same data on a log-log plot.\textbf{(b,c)} Resistance vs. scaled coupling energy for array H1 and H2. The black solid lines are the exponential fits across the critical point.}
 \end{figure*}
A similar scaling analysis was also carried out for samples H1 and H2. The results are presented in Fig. 3. (No scaling analysis was attempted in \cite{HavilandArrays}, so Fig. 2b and 2c are the original contribution of the present work). For these graphs, we found $z\approx1$, and $\nu\approx0.4$ in H1 and $\nu\approx0.5$ for H2. Several curves in these two graphs have tails that do not follow the common scaling behavior. These deviations correspond to the lowest temperatures and, as Fig. 1 attests, are in fact expected.

For all three samples, the scaled data close to the critical points can be well approximated by the exponential dependence, $y=a_1exp\left(-a_2x\right)$, shown in the figure with solid lines. Let's notice that the "scaling collapse" by itself can be coincidental \cite{Rogachev_Deficiency} so it is important to establish if there is any relation to the microscopic physics of the system. The next step of the analysis is to adapt the generic scaling Eq. 1 to the specific mechanism of a QPT in 1d JJ arrays and test if it fits the data. 

At strong coupling between the islands, the superconducting condensate in these arrays form a Tomonoga-Luttinger liquid state \cite{FazioLuttinger}. The JJ arrays act as a transmission line with capacitance $C_0$ to the ground and Josephson inductance $L_J$ (both per unit length) giving the propagation velocity $v_p=1/\sqrt{L_JC_0}$.  We are most concerned with the critical behavior of the chain, so we take both $L_J$ and $v_p$ at the critical point. From the parameters of the arrays we found that, for array B1, the velocity is $7.6\times{10}^5\ m/s$ and, for arrays H1 and H2, it is $3.6\times{10}^5$ m/s. Comparing our results to those found in \cite{KuzminManucharyan}, ($v_p=1.88\times{10}^6$ m/s) indicates that we are in a similar regime in terms of the wave environment. 

According to the general conjecture of our model, the dephasing length at the QPT in an interacting system is determined by the distance travelled by a system-specific excitation over the Planckian time.  Adapting this rule for the specific case of the superconducting plasmons we get  $L_\varphi\left(T\right)=v_p\tau_P=\left(1/\sqrt{L_JC_0}\right)\left(\hbar/k_BT\right)$.

With this choice of the dephasing length and continuing using $K_0$ as the driving parameter, the scaling equation takes the form
\begin{equation}
\frac{R}{T^{1/z}}=A\exp{\left(\frac{1}{T^{1/\nu z}}\frac{K_0-K_0^\ast}{K_0^\ast}\ \left(\frac{\hbar}{L_0k_B\left(L_J^\ast C_0\right)^{1/2}}\right)^{1/\nu}\right)}
\end{equation}

Using the parameter $a_2$ extracted from the exponential fit to the experimental data, $y=a_1exp\left(-a_2x\right)$, and the speed of the plasmons at the critical point, we determined the experimental value of the microscopic scale for each array, $L_0$. The values are given in Table 1.

\begin{table*}\caption{Relevant parameters for all arrays, listed here: array length, $N$; unit cell, $a$; screening length $\Lambda$; seeding length, $L_0$; dephasing length (calculated at $T=1K$), $L_\varphi$; Josephson inductance, $L_J$ (at the critical point) ; capacitance to ground, $C_0$; nearest-neighbor capacitance, $C_1$; correlation length critical exponent, $\nu$; and critical parameter $K_0^\ast$. $N$, $\Lambda$, $L_0$, and $L_\varphi$ are presented as a number of unit cells. $L_J$, $C_0$, and $C_1$ are calculated per unit cell.}
\smallskip

\begin{tabular}{l|ccccccccccc}

\hline
    &$N$  &$a$($\mu$m)  &$\Lambda$  &$L_0$  &$L_\varphi$($T$=1 K)   &$L_J^\ast$(nH) &$C_0$(fF)  &$C_1$(fF)  &$z\nu$  & $K_0^\ast$ &$E_J^*/E_1$
    \\[0.5ex]
  \hline
H1  &63     &0.2    &10     &10     &13.8   &7.18   &0.035  &3.5    &0.4    &0.10   &       \\
H2  &255    &0.2    &10     &10     &13.8   &8.75   &0.035  &3.5    &0.5    &0.10   &  \\
B1  &100    &1      &3.3    &3.4    &5.81   &6.40   &0.27   &3.0    &0.65   &0.30   &\\
  \hline
N1  &100x100&6      &11     &11     &9.47   &46.1   &0.014  &1.7    &0.9    &       &0.078    \\
M1  &190x60 &1      &11     &13     &15.8   &19.2   &0.012  &1.1    &0.6    &       &0.12     \\
\hline
\end{tabular}
\end{table*}

To provide experimental verification and check the self-consistency of our  "interacting plasmons" picture of the QPT, we need to determine an expected value for the seeding length, $L_0$. In the superconducting state, the plasmon is carried between islands by coherent tunneling of Cooper pairs. As the system becomes insulating, these Cooper pairs start to scatter off the junction barrier, causing a phase difference between the neighboring islands. Formal analysis of the Sine-Gordon model carried out in \cite{WuPhaseModes} demonstrates that these “opaque’ junctions contain an excess Cooper pair and, due to charge-phase duality, we can either treat these as domain walls with $\Delta\phi=\pm\pi$ or as localized charges on the islands with $q=\pm2e$.  A Cooper pair located on an island polarizes an array over the distance defined by the Coulomb screening length, $\Lambda=\sqrt{C_1/C_0}$. So, the screening length appears as a natural minimal length scale defining quantum states in a 1d JJ array. This assertion is supported by the experimental \cite{HavilandArrays,HavilandReview,HavilandSoliton} studies on JJ arrays with $C_1\gg C_0$ , where a soliton-like charge profile with width $\Lambda$ was observed.

A complimentary way to view $\Lambda$ as a minimal scale of the superconducting state is through the energy dispersion relation of the plasmons, given in \cite{MirlinSIT} as
\begin{equation}
\epsilon\left(k\right)=\frac{{\hbar\ \omega}_p\left|k\right|}{\sqrt{k^2+1/\Lambda^2}}
\end{equation}
where $\omega_p=\sqrt{E_JE_1}$ is the plasma frequency of a single junction. This dispersion relation effectively splits the plasmon behavior into two branches, separated by characteristic wavelength $\Lambda$. Plasmons with a larger wavelength (small k) will have a linear dispersion relation, while for shorter wavelength the spectrum is dispersionless. In other words, $\Lambda$ acts as the minimum wavelength of an excitation that will propagate while a shorter wavelength oscillation will be stationary. This picture of the length scale in the insulating regime is also consistent with the high-frequency, short-distance plasmons seen in the insulating regime in \cite{KuzminManucharyan}.

Using the numerical values for capacitances $C_0$ and $C_1$, we estimated the screening length for each array. For the 1d arrays, the numerical values for $\Lambda$ listed in Table 1 show remarkable agreement with the minimal scale $L_0$ extracted from the experiment. This finding confirms the microscopic picture of a QPT arising from interacting plasmons and is the main result of our paper.
 
Let us first of all note that the exact match between $L_0$ and $\Lambda$ is probably fortuitous; in general we expect agreement only to within a coefficient of order one. Let us further note that our analysis of the 1d arrays based on Eq. 3 is self-consistent. The obtained minimal scale is smaller than the dephasing length, which, in the range of the scaling analysis, $0.5\ <T<1\ K$, is in turn smaller than the length of the arrays. (The shortest $L_\varphi$ corresponding to $T=1\ K$ is listed in the Table). In other words, $L_0<L_\varphi<L$. These relations between the length scales are prerequisites for the system to undergo a QPT governed by the propagation and interaction of 1d plasmons.

Our model is phenomenological and so it does not address important questions such as what determines the critical point of the transition, the correlation exponent $\nu$, and the deviations from the scaling behavior at low temperatures. Let us now discuss these questions.

For the SIT in a long, 1d JJ array, the theoretically predicted critical parameter is $K_0=\sqrt{E_J/E_0}$ \cite{GlazmanLarkin}. Parameters $K_0$  and $E_0$ are defined slightly differently in different studies. The form used here follows reference \cite{BaskoQMC}, representing $K_0$ as the direct ratio of the Josephson coupling energy to the capacitive charging energy of a Cooper pair, instead of a single electron. With this definition, in the limit of a long array, the critical $K_0^\ast$ ranges from $3/\pi$ in the limit $C_0\gg C_1$ to $2/\pi$ when $C_1\gg C_0$. From the experiments we found the values of $K_0^\ast\approx0.1$ in arrays H1 and H2 and $K_0^\ast\approx0.3$ for array B1, both well below the theoretically predicted values. 

Looking at the table we can see that the transition actually occurs when $K_0^\ast\approx1/\Lambda$, or alternatively $K_1^\ast\approx1$ in all three arrays studied.  In view of the inconsistency in $K_0^\ast$, it is quite tempting to claim $K_1$ as the effective critical parameter.

If, on the other hand, one continues to insist that the transition is driven by $K_0$, our finding appears to be qualitatively similar to what was found in the microwave measurements in very long arrays \cite{KuzminManucharyan}. In these experiments, the superconducting response was found in samples that were expected to be on the insulating side of the transition. Similarly in DC experiments on short SQUID arrays, we find superconducting behavior in the range, $0.1<K_0<2/\pi$, where the arrays are expected to be insulating.

In the microwave studies of the long arrays it was proposed that the superconducting state with local correlations appears as the result of “melting” of the global insulating state \cite{HigginbothamMelting}. This work also provides an estimate for this melting temperature as $T_{ins}\approx\sqrt{2E_JE_1}/k_B\Lambda$. It is interesting to see if this temperature plays any role in the behavior of short SQUID arrays. Using for $E_J$ its critical value $E_J^\ast$, we found this “melting” temperature to be approximately $0.52\ K$ for array B1 and $0.42\ K$ for arrays H1 and H2. Comparing these temperatures to Fig. 1, we see that the superconducting behavior (lower branch of the plot) only occurs above $T_{ins}$, in an accurate agreement with the “melted” insulator picture. Furthermore, Ref. \cite{HigginbothamMelting} suggested that in this high-temperature regime the system is governed by the local superfluid stiffness $K_1$, which matches what we found for the three SQUID arrays.

Let us also mention one distinction between the studies. In the long arrays the superconductor-from-melted-insulator state is claimed to appear when $L_\varphi<\Lambda$ \cite{HigginbothamMelting}. While, in the short SQUID arrays in this paper, the system is always in the range $L_\varphi>\Lambda$, although the two lengths are quite close to each other.

The value of the correlation length exponent extracted from the experiment is in the range $\nu\approx0.4-0.65$; the exponent also appears to grow with the effective length of arrays $N/\Lambda$. For a very long array, the transition is discontinuous, which corresponds to $\nu=\infty$ \cite{HerbutNu}. The arrays studied here are much too short to approach this regime, instead there is apparently an observable transitional regime which can be fit with an effective exponent.  We find it instructive to compare this behavior with a recent Monte Carlo computation of the superfluid density, $\rho_s$, variation across the Bose-Hubbard transition in 1d \cite{BoseHubbard}. While for an infinite array $\rho_s$ undergoes a discontinuous jump, the finite-size arrays display continuous variation across the transition, which becomes less and less steep with decreasing $N$.

\begin{figure*}[tbph]
\centering
 \includegraphics[width= .75\textwidth]{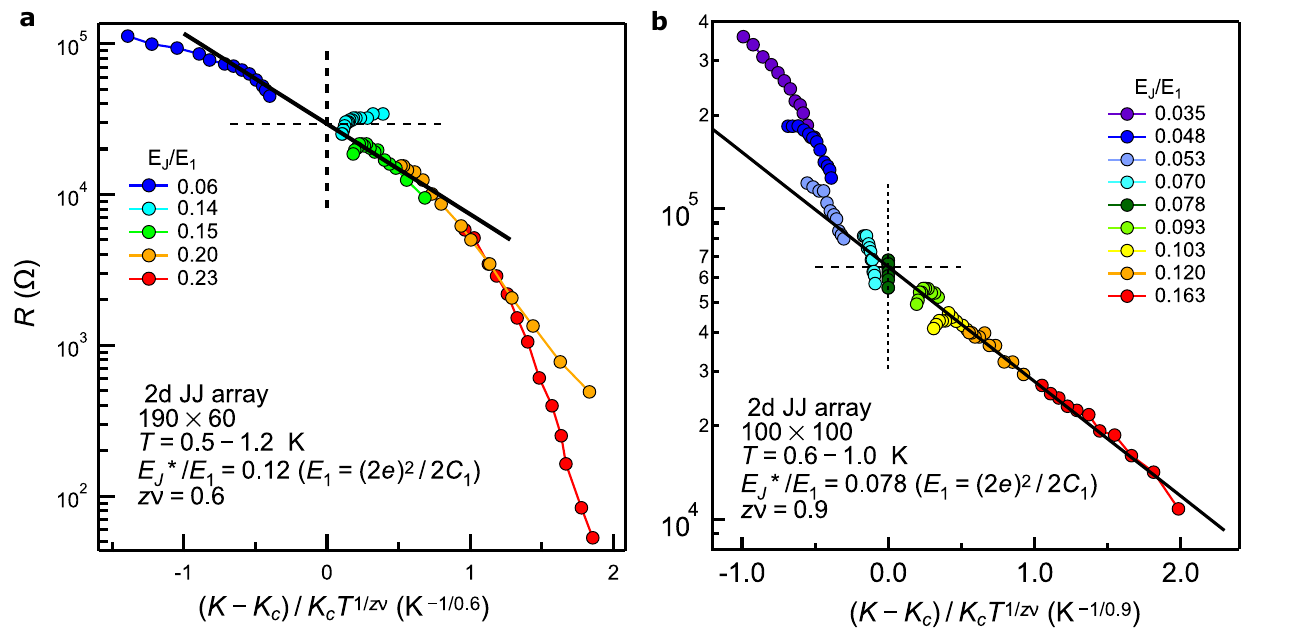}
 \caption{\textbf{Scaling analysis for 2D arrays.} 
 Resistance versus scaled coupling energy for (a) array M1 studied in \cite{ZantMooij2d} and (b) for array N1 studied in \cite{Nakamura2d}.}
 \end{figure*}
 
 \smallbreak
 \noindent \textbf{3. QPT in 2D array}
 
After seeing the success of our model in 1d, we decided to attempt to scale and analyze data of 2d JJ arrays in the same manner. Similar to the 1d case, a magnetic field was used to drive QPTs in several studies of 2d arrays \cite{Haviland2d,ZantMooij2d}. We have tried to analyze these data, but could not produce a scaling collapse, presumably due to complications caused by a frustration pattern with multiple QPTs. Fortunately, the authors of \cite{ZantMooij2d} also investigated the transition in a 190(long)$\times$60(wide) junction array in zero magnetic field by systematically varying $E_J$ in the fabrication process. We refer to this array as M1. Another very recent work on the QPT in 2d arrays studied the transition in 100$\times$100 square arrays \cite{Nakamura2d} in the same manner. We refer to this array as N1. 

The scaling analysis has not been presented in either of these studies, so we carry it out here.  The analysis in 2d is overall very similar to in 1d, but there are several differences. First, in 2d the prefactor in the scaling equation is temperature-independent. The downside to this is that we can not separate the critical exponents; instead we can only obtain their product, $z\nu$. Second, the driving parameter in 2d is predicted to be $E_J/E_1$ \cite{FazioVortex} for all arrays regardless of $\Lambda$. We also want to note a choice made in the definition of the charging energies, $E_0$ and $E_1$. For insulating arrays, these are usually defined as $e^2/2C$, based on the argument that the quasiparticles in the insulating state are single electrons \cite{FazioVortex}. We will continue to define them as $(2e)^2/2C$ to avoid confusion with our 1d analysis. Because the driving parameter is normalized, this will only affect the location of the critical point. $L_J$ for array M1 was calculated from $E_J$ values given in \cite{ZantMooij2d}.  For array N1, we used the ratio of $E_J/E_1$ and the capacitance values of $C_1\approx1.7$ fF and $C_0\approx14$ aF given in \cite{Nakamura2d} to find $L_J$ for each array. These and other parameters of M1 and N1 arrays are added to Table 1.  

Scaled data for arrays M1 and N1 are presented in Fig. 3. Similar to the 1d arrays, the studied data are limited to high temperatures since, below about 0.5 K, the data start to display reentrant behavior. All indicated values of $E_J/E_1$ are equivalent to those given in \cite{ZantMooij2d,Nakamura2d} but converted to $E_J/E_1$ and to the 2-electron form of $E_1$. The quality of the scaling collapse is not as good as in Fig. 2, but one should keep in mind that these figures combine data from different samples and hence are prone to unsystematic variation in the fabrication procedure and accuracy with which the parameter $E_J/E_1$ is determined. 

For sample M1, we found the experimental critical value of $E_J^*/E_1\approx0.12$ to be notably larger than the theoretical critical value of $2/\pi^2\approx0.2$ (0.051 in our notation) \cite{FazioVortex}. Assuming that $z=1$ as explained below, we found the critical exponent $\nu\approx0.6$. This is slightly lower than the predicted value from the XY model of $0.67$ \cite{FazioVortex}. For array N1 the scaled data are presented in Fig 3b. The critical value of $E_J^*/E_1\approx0.078$ was chosen to coincide with the value found in \cite{Nakamura2d} of (in their notation) $\approx0.31$. This is still higher than the expected critical value, but is much closer than for array M1. We found the critical exponent $\nu\approx0.9$, which is higher than the predicted value.

We are now in a position to test if the picture of interacting plasmons can account for the QPT in a 2d array. We use for this purpose Eq. 3 with two modifications: the prefactor for the 2d case is $T$-independent and the driving parameter is $E_J/E_1$ instead of $K_0$. Using the size of the unit cell, and the values of $C_0$ and $L_J^*$ given in Table 1, we found the speed of plasmons at the critical point, again using $v_p=a/(C_0 L_J^*)^{1/2}$ to be $2.1\times10^6$ m/s in sample M1 and $7.5\times10^6$ m/s in ssample N1. From the exponential fit to the data, $y=a_1 \exp (-a_2 x)$, shown as a solid line in Fig. 3a, we found $a_2$ for each system and then determined the microscopic scale of the QPT. For array M1, we found $L_0 \approx 13$ and in N1 we found it to be $\approx11$. Just as in the 1d case, this value is in remarkable agreement with the measured value of the screening length of both arrays $\Lambda\approx11$ \cite{ZantMooij2d,Nakamura2d}. 

Our observation confirms that we have a (perhaps effective) QPT in a small 2d array of Joshephson junctions and that near the critical point it is represented by the physics of interacting superconducting plasmons. This picture appears somewhat surprising to us since we expected the BKT vortex-anivortex physics of the superconducting phase to show up in some way in the quantum critical regime. 

\noindent \textbf{4. Summary. Comparison with the pair-breaking QPT in superconducting films and nanowires}

In summary, we found that a quantum phase transition takes place in three short 1d SQUID arrays and two small 2d arrays. In all five systems, the critical fluctuations are interacting plasmon modes and the microscopic seeding scale of the QPT is determined by the screening length $\Lambda$. These QPTs should probably be considered as effective ones, since the quantitative scaling analysis describes the data only at high temperatures above about 0.5 K. For 1d arrays, this temperature is consistent with the melting temperature of the insulating state by superconducting fluctuations proposed in Ref. \cite{HigginbothamMelting}. Moreover, for these 1d arrays, the QPT occurs deep in the insulating range as defined by the current theories. 

Let us now compare the QPT in short JJ arrays discussed in this paper with the magnetic-field-driven pair-breaking QPT (pbQPT) \cite{DelMaestroNanowire} observed in superconducting nanowires and films \cite{RogPB}. This comparison is especially interesting as both systems have very similar scaling behavior in 1 and 2 dimensions. 

In JJ arrays all resistance comes from superconducting fluctuations. In contrast, near the pbQPT superconductivity becomes gapless, and the first step of the analysis is to estimate and subtract the noncritical, dominant contribution of normal electrons. 

In JJ arrays, the critical fluctuations are "phase-only" excitations, superconducting plasmons propagating with linear dispersion. In pbQPT the critical fluctuations are of Aslamazov-Larkin type, which involve both amplitude and phase.

In both cases there is interaction between the fluctuations. Also, in the quantum critical regime of both scenarios, all effects of interactions are absorbed in the Planckian time. However, because of the different microscopic physics, the dephasing lengths of the two systems are different. In the arrays it is given by a "ballistic" expression $L_{\varphi}=v_p \hbar/ k_B T$ which implies that $z=1$. In nanowires and films, $L_{\varphi}=\sqrt{\hbar D/ k_B T}$, which corresponds to a diffusion of superconducting fluctuations through regions of normal metal and implies $z=2$. 

The experimental seeding scale $L_0$ emerges from very distinct microscopic physics and expectedly is very different. In the arrays it is given by the screening length $\Lambda=\sqrt{C_1/C_0}$ and in nanowires and films by the zero-temperature Ginzburg-Landau coherence length $\xi(0)$.   

Finally let us comment that our model is phenomenological and is not a replacement of a complete critical theory which starts from microscopic physics and goes all the way to the prediction of the long-range critical scaling behavior. For 2d superconducting systems, such description has not yet been developed. We hope it will be stimulated and assisted by our work.  It would also be interesting to see (for apparently quite a low computational cost, at least for 1d arrays) if the effective QPT observed in experiments is reproducible in the numerical procedures of Ref. \cite{MirlinSIT}, which incorporate both temperature and array length effects.
      
\smallbreak      
\noindent \textbf{ Acknowledgements.}
The authors thank D.M. Basko and A.D. Mirlin for useful comments on their work and W. Kuo for providing an extensive description of his 1d JJ arrays. This work was supported by the National Science Foundation under awards DMR1904221 and DMR2133014.

\end{document}